\documentclass[prd,superscriptaddress,twocolumn]{revtex4}

\usepackage{mathrsfs,amsmath}
\usepackage{eurosym}
\usepackage{amssymb}
\usepackage{bbold}

\usepackage{latexsym,epsfig,epstopdf}
\usepackage{amssymb,amsmath,amsthm}

\usepackage{times}
\usepackage{color}

\usepackage{lmodern}
\usepackage[utf8]{inputenc}

\usepackage{graphicx}
\usepackage{color}
\usepackage{braket}
\usepackage[mathscr]{euscript} 
\usepackage{perpage} 
\usepackage{slashed}

\usepackage{enumitem}

\def\be{\begin{equation}}
\def\ee{\end{equation}}
\def\bea{\begin{eqnarray}}
\def\eea{\end{eqnarray}}

\usepackage{calc}
\newsavebox\CBox
\newcommand\hcancel[2][0.5pt]{%
  \ifmmode\sbox\CBox{$#2$}\else\sbox\CBox{#2}\fi%
  \makebox[0pt][l]{\usebox\CBox}%
  \rule[0.5\ht\CBox-#1/2]{\wd\CBox}{#1}}

\begin{document}

\title{A solution to the soccer ball problem for generalised uncertainty relations} 

\author{Matthew J. Lake}
\email{matthew5@mail.sysu.edu.cn}
\affiliation{School of Physics, Sun Yat-Sen University, Guangzhou 510275, China}

\begin{abstract}

We propose a new method for generating generalised uncertainty relations (GURs) including the generalised uncertainty principle (GUP), extended uncertainty principle (EUP), and extended generalised uncertainty principle (EGUP), previously proposed in the quantum gravity literature, without modifying the Heisenberg algebra. 
Our approach is compatible with the equivalence principle, and with local Poincar{\' e} invariance in the relativistic limit, thus circumventing many of the problems associated with GURs derived from modified commutation relations. 
In particular, it does not require the existence of a nonlinear additional law  for momenta. 
This allows sensible multiparticle states to be constructed in which the total momentum is macroscopic, even if the momentum of an individual particle is bounded by the Planck momentum, thus providing a resolution of the ``soccer ball problem'' that plagues current approaches to GURs. 

\end{abstract}


\maketitle

\section{Introduction}

Gedanken experiments imply the existence of GURs in the quantum gravity regime, including the GUP, EUP and EGUP \cite{Scardigli:1999jh,Park:2007az}. 
However, in the absence of a complete quantum gravity theory, it remains an open problem how to derive such relations, more rigorously, from an underlying quantum formalism. 
To date, most attempts invoke modified commutation relations, but these remain plagued by theoretical difficulties, including the so-called ``soccer ball problem'' \cite{Hossenfelder:2014ifa} and violation of the equivalence principle (EP) \cite{Tawfik:2014zca}. 

Here, we propose an alternative scheme in which vectors in an infinite-dimensional Hilbert space, $|g_{\vec{x}}\rangle$, are associated with points, $\vec{x}$, in the classical background. 
This allows us to construct quantum superpositions of geometries, as expected in any candidate theory of quantum gravity. 
We then show that the probability distribution associated with the total quantum state, including contributions from canonical quantum matter and the quantum state of the geometry, naturally gives rise to both the GUP and the EUP. 
These may then be combined to give the EGUP. 

Crucially, in this approach, the Heisenberg algebra is unaffected apart from a rescaling of the form $\hbar \rightarrow \hbar + \beta$, where $\beta \simeq \hbar \times 10^{-61}$ is a function of $G$, $c$, $\hbar$ and $\Lambda$.  
Thus, we show that higher-order dispersion relations and / or modifications of Euclidean symmetry, which lead to modifications of the canonical position-momentum commutator, are {\it not} required to implement GURs \cite{Lake:2018zeg}. 
Our approach therefore avoids common problems associated with GURs derived from modified commutators, including both the soccer ball problem and violation of the EP. 
(Strictly, these problems are not solved, but circumvented, since they never arise in the first place.)

\section{Generalised uncertainty relations (GURs)}

\subsection{Recap of Heisenberg's uncertainty principle (HUP)}

The HUP contains the essence of canonical quantum mechanics (QM) and is a fundamental consequence of wave-particle duality. 
It can be introduced heuristically using the Heisenberg microscope thought experiment \cite{Rae}, yielding 
\begin{eqnarray} \label{HUP-heursistic}
\Delta x \, \Delta p \gtrsim \hbar \, , 
\end{eqnarray}
or derived from the quantum formalism using the position-momentum commutator, 
\begin{eqnarray}
[\hat{x},\hat{p}] = i\hbar \ \hat{\openone} \, , 
\end{eqnarray}
together with the Schr{\" o}dinger-Robertson relation \cite{Isham},
\begin{eqnarray} \label{Schrodinger-Robertson}
\Delta_\psi O_1 \, \Delta_\psi O_2 \geq \frac{1}{2} |\langle\psi | [\hat{O}_1,\hat{O}_2] | \psi\rangle| \, , 
\end{eqnarray}
where $\Delta_\psi O := \sqrt{ \langle\psi | \hat{O}^2 | \psi\rangle - \langle\psi | \hat{O} | \psi\rangle^2}$, giving
\begin{eqnarray} \label{HUP}
\Delta_\psi x \, \Delta_\psi p \geq \frac{\hbar}{2} \, .
\end{eqnarray} 
Note that, here, $\Delta_\psi x$ and $\Delta_\psi p$ are well defined, as standard deviations of the probability distribution ${\rm d}P(x|\psi) = |\psi|^2{\rm d}x$, as opposed to $\Delta x$ and $\Delta p$ in the heuristic example (\ref{HUP-heursistic}).

\subsection{Types of GUR and their physical interpretations: GUP, EUP and EGUP}

The standard GUP takes the form 
\begin{eqnarray} \label{GUP-heursistic}
\Delta x \gtrsim \frac{\hbar}{2\Delta p} + \alpha \frac{G}{c^3}\Delta p \, ,
\end{eqnarray} 
where $\alpha$ is a numerical constant, assumed to be of order unity. 
It can be derived heuristically by modifying the Heisenberg thought experiment to include the gravitational interaction between the massive particle and the probing photon \cite{Scardigli:1999jh}. 
This implies the existence of a minimum length scale, of the order of the Planck length, $l_{\rm Pl} := \sqrt{\hbar G/c^3} \simeq 10^{-33} \ {\rm cm}$. 

The EUP takes the form 
\begin{eqnarray} \label{EUP-heursistic}
\Delta p \gtrsim \frac{\hbar}{2\Delta x} + \eta \hbar\Lambda \Delta x \, .
\end{eqnarray} 
Here, $\Lambda \simeq 10^{-56} \ {\rm cm}$ is the cosmological constant \cite{Hobson:2006se} and $\eta$ is a numerical constant, also assumed to be of order one. 
Equation (\ref{EUP-heursistic}) can be obtained, heuristically, by considering the Heisenberg microscope experiment in the presence of dark energy: that is, by assuming the existence of a minimum spatial curvature of order $\sim \Lambda$, as opposed to asymptotically flat space \cite{Park:2007az}. 
This implies the existence of a minimum momentum scale, of the order of the de Sitter momentum $\sim m_{\rm dS}c$, where $m_{\rm dS} := (\hbar/c)\sqrt{\Lambda/3}$. 

Using either the GUP or EUP alone breaks position-momentum symmetry in the uncertainty relations, but both may be derived as separate limits of the EGUP, 
\begin{eqnarray} \label{EGUP-heursistic}
\Delta x \, \Delta p \gtrsim \frac{\hbar}{2} + \tilde{\alpha} (\Delta p)^2 + \tilde{\eta} (\Delta x)^2 \, , 
\end{eqnarray} 
which restores it \cite{Park:2007az}. 
Here, $\tilde{\alpha}$ and $\tilde{\eta}$ are appropriate dimensionful constants, which may be obtained by comparison with Eqs. (\ref{GUP-heursistic}) and (\ref{EUP-heursistic}). 
Thus, Eq. (\ref{EGUP-heursistic}) may be expected to combine the effects of both canonical (Newtonian) gravity and repulsion due to dark energy. 

Clearly, a more rigorous version of the EGUP, with $\Delta x$ and $\Delta p$ replaced by $\Delta_\psi x$ and $\Delta_\psi p$, respectively, can be derived from the quantum formalism by introducing the modified commutation relation \cite{Kempf:1996ss}
\begin{eqnarray} \label{mod-comm}
[\hat{x},\hat{p}] = i\hbar\hat{\openone} + \tilde{\alpha} \hat{x}^2 + \tilde{\eta} \hat{p}^2 \, .
\end{eqnarray} 
Equation (\ref{EGUP-heursistic}) then follows directly from the Schr{\" o}dinger-Robertson relation (\ref{Schrodinger-Robertson}).

\subsection{The soccer ball problem and violation of the equivalence principle}

Recall that, in the position space representation, the momentum operator may be identified with the shift-isometry generator of Euclidean space, up to a factor of $\hbar$:
\begin{eqnarray} \label{}
\hat{p} = -i\hbar \frac{\partial}{\partial x} =: \hbar\hat{k} =: \hbar \hat{d}_{x} \, . 
\end{eqnarray} 
Similarly, in the momentum space representation, the position operator may be identified with the shift-isometry generator of Euclidean momentum space (up to $\hbar$). 
Thus, the Heisenberg algebra is the algebra of the shift-isometry subgroup of the Galilean symmetry group. 
It may be obtained, rigorously, by combining Euclidean symmetries with the de Broglie relations, 
\begin{eqnarray} \label{}
\vec{p} = \hbar \vec{k} \, , \quad E = \hbar \omega \, , 
\end{eqnarray} 
or, equivalently, with the Hilbert space structure of canonical QM \cite{Isham}.
Therefore, modifying the canonical position-momentum commutator is equivalent to:
\begin{enumerate} 

\item  modifying the symmetry group of the classical background space 

\item introducing higher-order dispersion relations for quantum matter waves, i.e., the Fourier modes of the wave function $\psi(x)$, or 

\item both. 

\end{enumerate}

However, it is straightforward to show that the first possibility implies violation of Poincar{\' e} symmetry in the relativistic limit, leading to the soccer ball problem for multi-particle states \cite{Hossenfelder:2014ifa}. 
Furthermore, any deformation of the canonical commutator implies the existence of a mass-dependent gravitational acceleration, thus violating the equivalence principle. 
This is true regardless of whether such deformations arise from 1 or 2, or 3 (both) \cite{Tawfik:2014zca}.

Thus, the price paid to obtain the ``correct'' quantum gravity phenomenology using modified commutation relations, i.e., that expected from heuristic model-independent arguments, is extremely high: we are required to 
violate of founding principle of classical gravity! 

Theoretically, the EP may indeed be violated in the quantum gravity regime \cite{Hossenfelder:2012jw}. 
However, any such violation must be compatible with the construction of macroscopic multiparticle states, with the correspondence principle \cite{Rae}, and with the emergence of the classical world in the limit $\hbar \rightarrow 0$. 
This is clearly problematic for any canonical quantisation scheme based on modified commutators, due to the correspondence $\lim_{\hbar \rightarrow 0} [\hat{O}_1,\hat{O}_1]/(i\hbar) = \left\{O_1,O_2\right\}$. 
Below, we consider an alternative mechanism for generating GURs, based on quantum superpositions of geometries, which does not imply a significant modification of the Heisenberg algebra. 

\section{Quantum superpositions of geometries}

\subsection{Why do we need them?}

If quantum particles are to act as sources of the gravitational field, which is described by space-time curvature (i.e. geometry), then superpositions of position eigenstates should give rise to superpositions of geometries. 
In short, combining the principles of general relativity, including gravity as space-time curvature, and the principles of quantum mechanics, including the principle of quantum superposition, implies the existence of space-time superpositions. 
In the non-relativistic limit, it is reasonable to expect that these can be approximated by superpositions of spatial geometries, with a common absolute time parameter.  

\subsection{How do we get them?}

Here, we provide a concrete model that realises such geometric superpositions by ``smearing" the classical background space \cite{Lake:2018zeg}. 
The basic idea is to replace each point ``$\vec{x}$'' in the classical geometry by a superposition of all points. 
For each classical point, we obtain one whole copy of the classical space, i.e., a superposition of geometries, as desired.

\subsubsection{The smearing map in position space}

Since, in canonical QM, the classical point ``$\vec{x}$'' may be identified (heuristically) with a Dirac delta, or, equivalently, the ket $|\vec{x}\rangle$, this is most naturally realised by the map
\begin{eqnarray} \label{smearing-map}
|\vec{x}\rangle \mapsto |\vec{x}\rangle \otimes | g_{\vec{x}}\rangle \, ,
\end{eqnarray} 
where
\begin{eqnarray} \label{}
|g_{\vec{x}}\rangle := \int g(\vec{x}{\, '}-\vec{x}) |\vec{x}{\, '}\rangle {\rm d}^{d}\vec{x}{\, '} \, .
\end{eqnarray} 
Here, $d$ is the number of spatial dimensions and $g(\vec{x}{\, '}-\vec{x})$ is any normalised function. 
For simplicity, however, we may imagine $g$ as a Gaussian. 

Thus, we can visualise the smeared geometry associated with a classical $d$-dimensional universe $(\vec{x})$ as a $2d$-dimensional hyper-plane in which each point $(\vec{x},\vec{x}{\, '})$ is associated with a complex number $g(\vec{x}{\, '}-\vec{x})$. 
The map from the classical to the quantum phase space is then as follows:
\begin{eqnarray} \label{}
\vec{x} \leftrightarrow |\vec{x}\rangle \, , \ \ {\rm d}^d\vec{x} \leftrightarrow |\vec{x}\rangle{\rm d}^d\vec{x} \, , \ \ ( \ . \ , \ . \ ) \leftrightarrow . \otimes .
\end{eqnarray} 

We interpret $g(\vec{x}{\, '}-\vec{x})$ as the quantum probability amplitude for the transition $\vec{x} \rightarrow \vec{x}{\, '}$ \cite{Lake:2018zeg}. 
Hence, if $|g(\vec{x})|^2$ is peaked at the origin, the most probable value is $\vec{x}{\, '} = \vec{x}$. 
However, transitions to values within one standard deviation, $\sigma_g$, are relatively likely, giving rise to quantum fluctuations of the background geometry. 
We naturally associate the standard deviation of $g$ with the Planck scale, $\sigma_g \simeq l_{\rm Pl}$.

The choice of smearing function $g(\vec{x}{\, '}-\vec{x})$ and the map (\ref{smearing-map}) uniquely define the position space representation of the smeared-space formalism. 
It follows that the canonical QM state, $|\psi \rangle = \int \psi(\vec{x})|\vec{x}\rangle{\rm d}^{d}\vec{x}$, is mapped according to $|\psi \rangle \rightarrow |\Psi \rangle$, where
\begin{eqnarray} \label{}
|\Psi\rangle := \int\int g(\vec{x}{\, '}-\vec{x}) \psi(\vec{x})|\vec{x}\rangle\otimes |\vec{x}{\, '}\rangle {\rm d}^d\vec{x}{\rm d}^d\vec{x}{\, '} \, . 
\end{eqnarray} 

Since, in the smeared geometry, an observed value ``$\vec{x}{\, '}$'' does not determine which point(s) underwent the transition $\vec{x} \rightarrow \vec{x}{\, '}$, we must sum over all possibilities by integrating the joint probability density $|\Psi(\vec{x},\vec{x}{\, '})|^2 := |g(\vec{x}{\, '}-\vec{x})|^2|\psi(\vec{x})|^2$ over ${\rm d}^{d}\vec{x}$, yielding
\begin{eqnarray} \label{Prob}
\frac{{\rm d}^{d}P(\vec{x}{\, '} | \Psi)}{{\rm d}\vec{x}{\, '}^{d}} = (|g|^2 * |\psi|^2)(\vec{x}{\, '})  \, . 
\end{eqnarray} 
Since the variance of a convolution is equal to the sum of the variances of the individual probability distributions, we then have
\begin{eqnarray} \label{X-GUR}
(\Delta_{\Psi}x'^{i})^2 = (\Delta_{\psi}x'^{i})^2 + (\Delta_g x'^{i})^2 \, .
\end{eqnarray} 

It is straightforward to show that the generalised position-measurement operator, $\hat{X}^{i}$, given by
\begin{eqnarray} \label{}
\hat{X}^{i} := \int x'^{i} |\vec{x}\rangle\langle\vec{x}| \otimes |\vec{x}{\, '}\rangle\langle\vec{x}{\, '}| {\rm d}^d\vec{x} {\rm d}^d\vec{x}{\, '} = \hat{\openone} \otimes \hat{x}'^{i} \, , 
\end{eqnarray} 
yields the correct statistics, i.e. $(\Delta_\Psi X^{i})^2 = \langle\Psi |(\hat{X}^{i})^{2}|\Psi\rangle - \langle\Psi|\hat{X}^{i}|\Psi\rangle^2
= (\Delta_\psi x'^{i})^2 + (\sigma_g^i)^2$, where $\sigma_g^i \equiv \Delta_g x'^{i}$. 
Setting $\sigma_g^i = l_{\rm Pl}$ for all $i$, using the HUP (\ref{HUP}), and Taylor expanding (\ref{X-GUR}) to first order, we then obtain the GUP:
\begin{eqnarray} \label{GUP*}
\Delta_\Psi X \gtrsim \frac{\hbar}{2 \Delta_\psi p'} + \frac{2G}{c^3} \Delta_\psi p' \, . 
\end{eqnarray} 

\subsubsection{The smearing map in momentum space}

In the momentum space representation, we define an exactly analogous formalism. 
Thus, the canonical QM state $|\psi \rangle = \int \tilde{\psi}(\vec{p})|\vec{p}\rangle{\rm d}^{d}\vec{p}$ maps to
\begin{eqnarray} \label{}
|\Psi\rangle := \int\int \tilde{g}_{\beta}(\vec{p}{ \, '}- \vec{p})\tilde{\psi}_{\hbar}(\vec{p})|\vec{p} \, \vec{p}{ \, '}\rangle {\rm d}^d\vec{p} {\rm d}^d\vec{p}{ \, '} \, ,
\end{eqnarray} 
where $\tilde{g}_{\beta}(\vec{p}{ \, '}- \vec{p})$ is interpreted as the quantum probability amplitude for the transition $\vec{p} \rightarrow \vec{p}{ \, '}$ in momentum space. 
(The meaning of the index $\beta$ will be made clear soon.) 
Here, $|\vec{p} \, \vec{p}{\, '}\rangle$ is a basis vector in the enlarged Hilbert space, labelled by the values $\vec{p}$ and $\vec{p}{\, '}$, but is {\it not} a simple tensor product state. 
Consistency then requires that
\begin{eqnarray} \label{}
\langle \vec{x}| \langle \vec{x}{ \, '} | \vec{p} \, \vec{p}{ \, '} \rangle := \frac{1}{2 \pi \sqrt{\hbar \beta}} e^{\frac{i}{\hbar} \vec{p}.\vec{x}} \, e^{\frac{i}{\beta}(\vec{p}{\, '}-\vec{p}).(\vec{x}'-\vec{x})} \, 
\end{eqnarray}
and 
\begin{eqnarray} \label{}
\tilde{g}_\beta(\vec{p}{ \, '} - \vec{p}) := \frac{1}{\sqrt{2 \pi \beta}} \int g(\vec{x}{ \, '}-\vec{x}) \, e^{- \frac{i}{\beta}(\vec{p}{ \, '}- \vec{p}).(\vec{x}{ \, '}-\vec{x})} {\rm d}^d\vec{x}{ \, '} 
\nonumber
\end{eqnarray}
hold, with $\beta \neq \hbar$, in addition to the usual relations $\langle \vec{x} | \vec{p} \rangle = \frac{1}{\sqrt{2\pi \hbar}} \, e^{\frac{i}{\hbar} \vec{p}.\vec{x}}$ and $\tilde \psi_{\hbar} (\vec{p}) =
\frac{1}{\sqrt{2 \pi \hbar}} \ \int \psi (\vec{x}) \, e^{-\frac{i}{\hbar} \vec{p}.\vec{x}} \, {\rm d}^d\vec{x}$ \cite{Lake:2018zeg}.

In canonical QM, the momentum space representation of the wave function is the Fourier transform of the position space representation, $\psi(\vec{x})$, transformed at the scale $\hbar$. 
This follows directly from the canonical de Broglie relations plus the basis-independence of $|\psi\rangle$. 
In the smeared-space formalism, $\tilde{g}_\beta(\vec{p}{ \, '} - \vec{p})$ is the Fourier transform of $g(\vec{x}{\, '}-\vec{x})$, transformed at the scale $\beta$. 
This follows directly from the basis independence of $|\Psi\rangle$, which implies modified de Broglie relations for matter waves propagating on the smeared-space background, $\vec{p}{\, '} = \hbar\vec{k} + \beta(\vec{k}{\, '}-\vec{k})$ \cite{Lake:2018zeg}. 

Thus, analogues of Eqs. (\ref{Prob})-(\ref{X-GUR}) also hold for the momentum space representation of the smeared-state $|\Psi\rangle$, so that, defining the generalised momentum operator as $\hat{P}_{i} := \int \int p'_{i} |\vec{p} \, \vec{p}{ \, '}\rangle \langle \vec{p} \, \vec{p}{\, '}| {\rm d}^d\vec{p} {\rm d}^d\vec{p}{ \, '}$, we immediately obtain $(\Delta_\Psi P^{i})_2 = \langle\Psi |(\hat{P}^{i})_{2}|\Psi\rangle - \langle\Psi|\hat{P}_{i}|\Psi\rangle^2 = (\Delta_\psi p'_{i})^2 + (\tilde{\sigma}_{gi})^2$, where $\tilde{\sigma}_{gi} \equiv \Delta_g p'^{i}$. 
Setting $\tilde{\sigma}_{gi} = (1/2)m_{\rm dS}c$ for all $i$, using the HUP (\ref{HUP}), and Taylor expanding to first order, we obtain the EUP:
\begin{eqnarray} \label{GUP*}
\Delta_\Psi P \gtrsim \frac{\hbar}{2 \Delta_\psi x'} + \frac{\hbar \Lambda}{12} \Delta_\psi x' \, . 
\end{eqnarray} 
The transformation scale for the smearing function $g$ can then be written as $\beta = (2/d)\sigma_g^i\tilde{\sigma}_{gi}$, so that, for $d=3$ (our observed universe), $\beta = 2\hbar\sqrt{\rho_{\Lambda}/\rho_{\rm Pl}} \simeq \hbar \times 10^{-61}$, where $\rho_{\rm Pl} := c^5/(\hbar G^2) \simeq 10^{93} \ {\rm g \, cm^{-3}}$ is the Planck density and $\rho_{\Lambda} := \Lambda c^2/(8\pi G) \simeq 10^{-30}  \ {\rm g \, cm^{-3}}$ is the observed dark energy density \cite{Lake:2018zeg}.

\subsection{Implications: resolution of the soccer ball problem and restoration of the equivalence principle} 

It is straightforward to show that in the smeared-space formalism (outlined above) the generalised position-momentum commutator takes the same basic form as in canonical QM, but with a tiny rescaling of Planck's constant such that $\hbar \rightarrow \hbar + \beta$, i.e., 
\begin{eqnarray} \label{HUP-rescaled}
[\hat{X}^{i},\hat{P}_{j}] = i(\hbar + \beta)\delta^{i}{}_{j} \, {\bf\hat{\openone}} \, . 
\end{eqnarray} 
Defining the Hamiltonian as $\hat{H} := P^2/(2m)$, where $P$ is the absolute value of the generalised momentum, and the smeared-space potential as $\hat{\mathcal{V}} := \hat{\openone} \otimes \hat{V}'$, by analogy with $\hat{X}^{i} := \hat{\openone} \otimes \hat{x}'^i$, we may construct the Heisenberg equation for the smeared-state $|\Psi\rangle$. 
This takes the same basic form as the canonical Heisenberg equation, but with $\hbar \rightarrow \hbar + \beta$. 
The equation of motion for $\hat{X}^i$ is then
\begin{eqnarray} \label{}
\frac{d\hat{X}^i}{dt} = -\frac{i}{(\hbar + \beta)}[\hat{X}^{i},\hat{P}_{j}]\frac{\partial\hat{H}}{\partial P_{j}} = \frac{\hat{P}^i}{m} \, , 
\end{eqnarray} 
so that the acceleration of the position expectation value is independent of the particle mass, as in canonical QM. 
Clearly, this is not the case for modified commutators of the form (\ref{mod-comm}). 
Hence, although such modifications yield both the GUP and the EUP, they are fundamentally incompatible with the EP. 
By contrast, the smeared-space formalism yields both the GUP and EUP without violating the EP. 

In \cite{Lake:2018zeg}, it was also shown how to construct multiparticle states in the smeared-space background. 
As a general operator $\hat{O}$ may be written as a function of the operators $\hat{X}^i$ and $\hat{P}_j$, such states are compatible with the correspondence $\lim_{\hbar + \beta \rightarrow 0} [\hat{O}_1,\hat{O}_1]/(i(\hbar+\beta)) = \left\{O_1,O_2\right\}$ in the macroscopic limit. 
By contrast, implementing a canonical quantisation scheme with modified commutation relations, and requiring the correspondence principle to hold, implies an equivalent modication of the canonical Poisson brackets. 
This implies violation of Galilean invariance, even for macroscopic systems, and, hence, violation of Poincar{\' e} invariance in the relativistic limit. 

Furthermore, if modifications of the canonical commutators are assumed to arise from nonlinear corrections to $p(k)$ in the relativistic regime, it is unclear whether one should require the physical momentum
$p$, or wave number $k$ (also known as the pseudo-momentum), to transform under the Poincar{\' e} group. 
However, in either case, the Lorentz transformations become nonlinear functions of the relevant quantity \cite{Hossenfelder:2012jw}. 
Thus, if the nonlinear momentum composition function has a maximum at the Planck momentum, corresponding to a minimum length of order $\sim l_{\rm Pl}$, the sum of momenta can never exceed this maximum value. 
It is therefore unclear whether mutliparticle states with macroscopic momenta can be constructed in models with modified commutation relations, and the problem of reproducing a sensible multiparticle limit is known as the soccer ball problem \cite{Hossenfelder:2014ifa}, as discussed above. 
In the smeared-space formalism, this problem does not arise, since we obtain GURs without modifying the fundamental symmetries of canonical QM and their associated algebras, i.e., commutation relations, except for the rescaling 
$\hbar \rightarrow \hbar + \beta$. 

\section{Conclusions}

We have proposed an alternative model of quantum gravity phenomenology in the non-relativistic regime in which GURs, including the GUP, EUP and EGUP, previously proposed in the quantum gravity literature, arise from quantum superpositions of the spatial background. 
Crucially, our approach leaves the commutation relations of canonical QM unchanged, except for a simple rescaling of the form $\hbar \rightarrow \hbar + \beta$. 
Thus, we have shown how GURs may be obtained within a well defined quantum formalism without assuming modified commutation relations, which are known to lead to theoretical difficulties including the soccer ball problem and violation of the EP. 

\section{Acknowledgements}

This work was supported by the Natural Science Foundation of Guangdong Province grant no. 2016A030313313 and the Singapore MOE Academic Research Fund Tier 1 project no. RG106/17.


\end{document}